\documentclass[iop,apj]{emulateapj}
\usepackage{natbib}
\usepackage{aas_macros}
\usepackage{xspace}
\usepackage{epstopdf}

\newcommand{\fullpath}{}
\newcommand{\cch}{C$_{2}$H\xspace}
\newcommand{\chcch}{CH$_{3}$CCH\xspace}
\newcommand{\hco}{HCO$^{+}$\xspace}
\newcommand{\hhco}{H$_{2}$CO\xspace}
\newcommand{\eqq}{\!=\!}  
\newcommand{\too}{\!\rightarrow\!} 
\newcommand{\jone}{{$J\eqq 1\too0$}}
\newcommand{\jtwo}{{$J\eqq2\too1$}}
\newcommand{\jthree}{{$J\eqq3\too2$}}
\newcommand{\jfour}{{$J\eqq4\too3$}}
\newcommand{\jfive}{{$J\eqq5\too4$}}
\newcommand{\jsix}{{$J\eqq6\too5$}}
\newcommand{\jseven}{{$J\eqq7\too6$}}

\newcommand{\jtwelve}{{$J\eqq12\too11$}}
\newcommand{\jthirteen}{{$J\eqq13\too12$}}
\newcommand{\jfourteen}{{$J\eqq14\too13$}}
\newcommand{\jfifteen}{{$J\eqq15\too14$}}
\newcommand{\jsixteen}{{$J\eqq16\too15$}}

\newcommand{\as}{$^{\prime\prime}$}
\newcommand{\mm}{$\rm \mu m$}

\newcommand{\zspecrange}{190--307~GHz\xspace}

\shorttitle{High-Density Molecular Gas in M82}
\shortauthors{Naylor et al.}
\begin{document}
\bibliographystyle{apj}
\title{A Census of the High-Density Molecular Gas in M82}




\author{B.J. Naylor\altaffilmark{1}, 
C.M. Bradford\altaffilmark{1},
J.E. Aguirre\altaffilmark{2},
J.J. Bock\altaffilmark{1},
L. Earle\altaffilmark{3},
J. Glenn\altaffilmark{3},
H. Inami\altaffilmark{4},
J. Kamenetzky\altaffilmark{3},
P.R. Maloney\altaffilmark{3},
H. Matsuhara\altaffilmark{4},
H.T. Nguyen\altaffilmark{1},
J. Zmuidzinas\altaffilmark{1}}
\altaffiltext{1}{Jet Propulsion Laboratory, California Institute of Technology, Pasadena, CA, 91109}
\altaffiltext{2} {University of Pennsylvania, Philadelphia, PA 19104}
\altaffiltext{3}{Dept. of Astrophysical and Planetary Sciences, University of Colorado, 389-UCB, Boulder, CO 80309}
\altaffiltext{4}{ISAS / JAXA, Sagamihara, Japan}

\slugcomment{Astrophysical Journal, in press}

\begin{abstract}
We present a three-pointing study of the molecular gas in the starburst nucleus of M82 based on \zspecrange spectra obtained with Z-Spec at the Caltech Submillimeter Observatory.   We present intensity measurements, detections and upper limits, for 20 transitions, including several new detections of CS, HNC, \cch, \hhco, and \chcch lines.   We combine our measurements with previously-published measurements at other frequencies for HCN, HNC, CS, C$^{34}$S, and HCO$^+$ in a multi-species likelihood analysis constraining gas mass, density and temperature, and the species' relative abundances.  We find some 1.7--2.7$\times10^{8}\, M_{\odot}$ of gas with $n_{\rm H_2}$ between 1--6$\times 10^4\,\rm cm^{-3}$ and $T>$50~K.  While the mass and temperature are comparable to values inferred from mid-$J$ CO transitions, the thermal pressure is a factor of 10--20 greater.  The molecular interstellar medium is largely fragmented and is subject to ultraviolet irradiation from the star clusters.  It is also likely subject to cosmic rays and mechanical energy input from the supernovae, and is warmer on average than the molecular gas in the massive star formation regions in the Milky Way.  The typical conditions in the dense gas in M82's central kpc appear unfavorable for further star formation;  if any appreciable stellar populations are currently forming, they are likely biased against low mass stars, producing a top-heavy initial mass function.
\end{abstract}


\section{Introduction}

Studies of molecular gas - the actively star-forming part of the
interstellar medium (ISM) - in other galaxies have been carried out most
extensively in the low-lying rotational transitions of the CO
molecule. Due to its high abundance relative to molecular hydrogen and
its favorable energy level spacing as compared to typical molecular cloud
temperatures, CO produces the brightest lines in
the millimeter-wavelength portion of the spectrum.

However, while these lines trace the bulk of the molecular gas, the
small dipole moment of CO results in modest critical densities for
these lines, $\sim 10^3 - 10^4$ cm$^{-3}$, and thus they do not
strongly discriminate between high-density star-forming cores and more
diffuse gas. High-dipole-moment molecules, such as HCN and CS, have
critical densities and radiative rates that are $100-1000$ times
larger than CO. Despite their much smaller abundances, these species
can be used to probe the dense ($10^4 - 10^7$ cm$^{-3}$) cloud cores
believed to be associated with star formation (SF).  Early measurement of HCN and CS in the Galaxy \citep{Morris:1974, Turner:1973} and in external galaxies \citep{Henkel:1988mz,Solomon:1992fj} showed that their intensities are well-correlated with the total far-infrared (FIR) flux.   More recently, HCN \jone\ luminosity has been shown to be directly proportional to FIR luminosity (a proxy for star formation rate [SFR]) in a sample of $\sim$100 normal spiral and starburst galaxies \citep{Gao:2004b,Gao:2004a}, as well as individual star-formation sites in the Galaxy \citep{Wu:2005}.

Insofar as HCN \jone\ measures dense gas mass, the $L_{\rm FIR}$ / HCN correlation across 7--8 orders of magnitude in luminosity implies a scale-independent relationship between dense gas mass and SFR.  An accurate assessment of the physical conditions in the HCN-emitting gas as well as the mass scaling ($M_{\rm H_2} / I_{\rm HCN}$) is thus of universal interest for theoretical SF studies.  Is the HCN-emitting gas simply a bi-product of star formation, perhaps formed in shocks or outflows and not participating in the formation of new stars, or is some of it the very material from which new stars form?  Such an assessment is best made with multiple transitions of HCN and by including transitions from other high-dipole-moment molecules where available.  

To assess the average properties of the dense gas on the scale of a nuclear starburst,  we have observed the nucleus of the M82 in a suite of millimeter-wave transitions of high-dipole moment species.  
The brightest infrared (IR) galaxy in the sky due to its proximity \citep[3.9 Mpc,][]{Sakai:1999lr}, M82 radiates an infrared luminosity \citep[$L=5.9 \times 10^{10} \,L_{\odot}$,][]{Sanders:2003bh}, exceeding that of the Galaxy, from a region that is only about 450 pc in radius \citep[e.g.,][]{Leeuw:2009qy}. Because of this concentration of star-forming activity, M82 has been dubbed the prototypical starburst galaxy, which makes it a particularly interesting laboratory for the study of SF.
It has been suggested for over 30 years that the stellar initial mass function (IMF) in M82 (and presumably other starburst nuclei) is biased against low-mass stars relative to the the local IMF \citep{Rieke:1980, Rieke:1993, Forster:2003}, but this has been debated \citep[e.g.,][]{Satyapal:1997,Colbert:1999dq}.  If the IMF is indeed low-mass deficient, a plausible line of inquiry is the initial conditions of SF -- the temperature and density of the dense molecular cloud cores.

We have obtained full \zspecrange spectra with the Z-Spec 1-mm grating spectrometer, which accesses the \jthree\ transitions of HCN, HCO$^+$, and HNC and \jfour, \jfive, and \jsix\ transitions of CS, with a uniform calibration.  Our study benefits from the prior observations of \jone\ transitions of HCN, HNC, HCO$^+$ \citep{Nguyen-Q-Rieu:1989pd,Huettemeister:1995uq}, as well as \jfour\ transitions of HCN, HCO$^+$ \citep{Seaquist:2000qf}.  These data are combined with our observations to 
generate the first comprehensive multi-species excitation and radiative transfer model for the dense gas in this source.   Of course, since M82 is also well-studied in multiple CO transitions \citep{Wild:1992qy,Mao:2000fr, Ward:2003ai, Weiss:2005, Seaquist:2006,Panuzzo:2010} as well as in the mid- and far-IR atomic gas tracers \citep{Forster:2003,Colbert:1999dq}, we have the opportunity to put the dense gas into context with the other ISM components, as well as the general properties of this prototypical nuclear starburst.

\section{Observations}

\begin{figure*}
\begin{center}
\includegraphics[width=\textwidth]{\fullpath{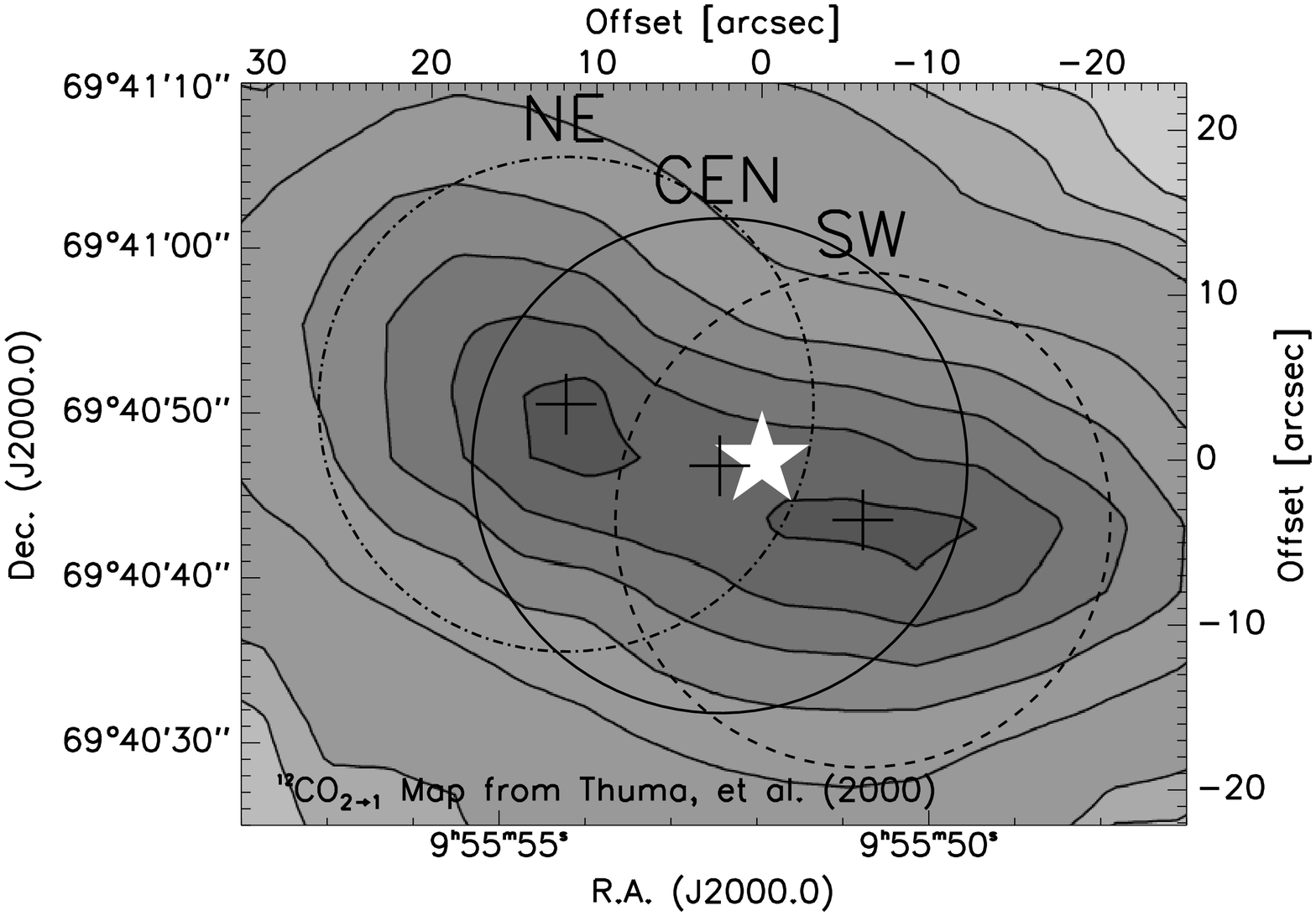}}
\caption[Pointings observed in M82]{Pointings toward M82. Z-Spec's 30\arcsec\ beam (FWHM at 250 GHz) is plotted over the single-dish measurements of the $^{12}$CO \jtwo\ map from \citet{Thuma:2000uq}.  The white star marks the 2.2 $\mu$m peak at $\alpha_{J2000.0}=9^\mathrm{h}55^\mathrm{m}51.9^\mathrm{s}$, $\delta_{J2000.0}=69^{\circ}40^{\prime}47.14^{\prime\prime}$ which is the adopted center for the observations.  The offsets relative to the center and integration times are given in Table \ref{tab:pointings}.  All three pointings use a 90\arcsec\ symmetric azimuthal chop throw.}
\label{fig:pointings}
\end{center}
\end{figure*}

Z-Spec is the first grating spectrometer for the millimeter band; it covers the full \zspecrange range instantaneously, dispersing this band to an array of 160 bolometers.  More information can be found in \citet{Glenn:2007qy}, \citet{Bradford:2009} and the Society of Photo-Optical Instrumentation Engineers (SPIE) articles: \citep{Naylor:2003,Bradford:2004,Earle:2006}.   The instrument operates at the Nasmyth focus of the Caltech Submillimeter Observatory (CSO) atop Mauna Kea.  While the instrument is $1/f$ stable down to $\sim$~100~mHz, we use a chop and nod mode to avoid the atmospheric fluctuations, which become important relative to the fundamental noise sources at $\sim$0.3--1~Hz, depending on the atmospheric conditions.   For the M82 observations the chop frequency was 1.6~Hz, the throw was 90\as\ in azimuth, and the nod interval was 20 seconds.  Three pointings along M82's major axis were observed on 2009 January 5 as summarized in Figure~\ref{fig:pointings} and Table~\ref{tab:pointings}.  The M82 spectra are calibrated using Mars, with an interpolation scheme using bolometer operating voltages as a measure of response \citep{Earle:2008uq, Bradford:2009}.  The data were reduced with standard demodulation and differencing appropriate for the chop and nod observing mode and Uranus is used as a spectral flat-fielder.  We expect the channel-to-channel calibration uncertainties to be less that 10\% except at the lowest frequencies which are extremely sensitive to the wing of the 186 GHz atmospheric water line.

\begin{deluxetable}{ccccc}
\tablecaption{M82 Observed Positions\label{tab:pointings}}
\tablewidth{0pt}
\tablehead{
\colhead{Pointing} & \colhead{$\alpha$~offset} & \colhead{$\delta$~offset} & 
\colhead{Int.~Time} & \colhead{Sensitivity} \\
 & \colhead{(arcsec)} & \colhead{(arcsec)} & \colhead{(min)} & \colhead{(Jy s$^{1/2}$)}
 }
\startdata
NE     & $+12.2$ &   $+3.3$ & 60.2 & 1.3 \\
CEN  & $+2.7$ &   $-0.5$ & 68.6 & 1.4    \\
SW    & $-6.1$ &   $-3.8$ & 60.1 & 1.2    
\enddata
\tablecomments{The R.~A.~and Dec.~offsets are relative to $\alpha_{J2000.0}=9^\mathrm{h}55^\mathrm{m}51.9^\mathrm{s}$, $\delta_{J2000.0}=69^{\circ}40^{\prime}47.14^{\prime\prime}$.  The integration time is the total demodulated time; the on-source time is half the listed values.  The quoted sensitivity is the median value of channel errors multiplied by the square root of the integration time and does not represent the ultimate sensitivity of the instrument; Z-Spec's sensitivity to spectral lines and to fainter sources is better by at least a factor of two.  The optical depth during these observations was $\tau_{225 \mathrm{GHz}} = 0.08-0.1$.}
\end{deluxetable}

\section{Results}

\begin{figure*}
\begin{center}
\includegraphics[width=\textwidth]{\fullpath{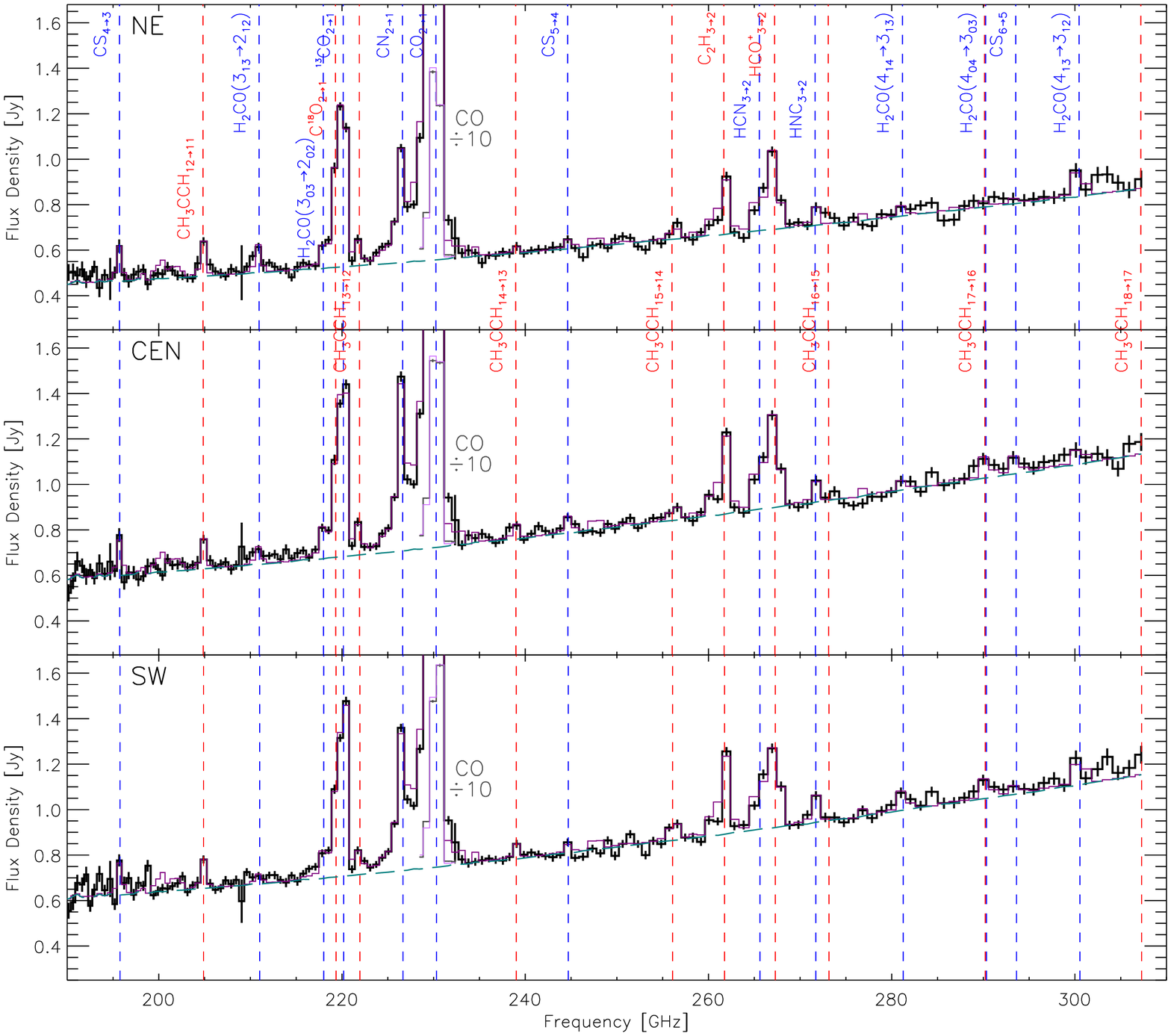}}
\caption[Spectra of three pointings]{Z-Spec \zspecrange spectra toward three positions in the nucleus of M82: NE at the top, CEN in the middle, and SW on the bottom.  The black histogram with error bars are the calibrated measurements and the purple histogram is the spectral fit, including both continuum and 20 fitted lines.  The fitted continuum is also plotted in dashed dark green and the fitted lines are marked with vertical blue and red dashed lines.  The dominant CO \jtwo\xspace line is shown in gray on the plot, scaled down by a factor of ten; the fit to the CO line, also scaled down, is shown in lavender.  The \chcch~\jsixteen~transitions and above are shown for reference but are not included in the fit.  We have identified all three sigma or greater spectral-line features that appear in all three pointings.  Nonetheless, there is additional structure in the spectra probably due to a multitude of blended, weak lines which Z-spec cannot individually identify.  The results from the fits are given in in Tables \ref{tab:contfit} and \ref{tab:lineinten}.}
\label{fig:ThreeSpec}
\end{center}
\end{figure*}

Spectra for the NE, CEN, and SW pointings are shown in Figure \ref{fig:ThreeSpec}.  The general features agree with previous measurements of the 1.2-mm continuum and CO~\jtwo , such as those in Figure~\ref{fig:pointings}.  

\subsection{Continuum Fluxes and Fits}

\label{ContFit}
The continuum flux in the Z-Spec band is due to a combination of the thermal dust emission which dominates at higher frequencies and the free-free emission from [H {\small II}] region electrons which dominates at lower frequencies.   The dust emission is parameterized as

\begin{equation}
\label{eqn:graybody}
F_{\mathrm{T}}(\nu) = \Omega B_{\nu} (T) 
	\left\{1-\exp\left[-\left(\frac{\lambda_{0}}{\lambda}\right)^{\beta}\right]\right\}, 
\end{equation}
where $B_{\nu} (T) = (2 h \nu^{3}/c^{2})/(\exp[h\nu/kT] - 1)$ is the Planck function.  \citet{Hughes:1994lr} fit a collection of observations from 3.3 mm to 40 $\mu$m and derive $\Omega = 1.34\times10^{-8}$ sr, $T = 48.1$ K, $\beta = 1.3$ and $\lambda_{0} = 7.9$ microns for the thermal dust.  The free-free emission is given by a simple power law,
\begin{equation}
\label{eqn:freefree}
F_{\mathrm{ff}}(\nu) = F_{0}\left(\frac{\nu}{\nu_{0}}\right)^{-0.1},
\end{equation}
where $F_{0}$ is the free-free flux at $\nu_{0}$.  Analysis of a map of M82 at 92 GHz found that the emission at that frequency is dominated by free-free emission.  \citet{Carlstrom:1991fk} calculate the free-free flux density $F_{\mathrm{ff}}$(92 GHz)$= 0.5 \pm 0.1$ Jy.  In the Z-Spec band, the free-free emission accounts for roughly one quarter of the total continuum flux, but that is significant enough that a dust-only fit to our spectra does not match the continuum level.  Instead, we model the continuum flux we observe to be a beam-scaled fraction of the total continuum emission of the galaxy,
\begin{equation}
\label{eqn:ContFit}
F(\nu) = A\left(\frac{\nu}{240 \,\mathrm{GHz}}\right)^{B-2}
	\left[F_{\mathrm{T}}(\nu) + F_{\mathrm{ff}}(\nu)\right].
\end{equation}
This model assumes that the free-free and thermal continuum emission have the same spatial distribution, which is reasonable given our coarse spatial resolution.  It also accounts for the spectral dependence caused by the spatial distribution.  If M82 completely filled our beam at all frequencies, we would expect $B = 0$, while $B = 2$ would be the prediction for a point source.  We find that $B = 1$ is the best fit value for all three pointings (see Table \ref{tab:contfit}), which makes sense given that M82 is observed edge on and is thus roughly point-like in one dimension and beam-filling in the other, relative to our 30\arcsec~beam.  The continuum fraction seen in the three pointings is consistent with the 1.2 mm continuum distribution shown in \citet{Thuma:2000uq}.  The peak of the 1.2 mm continuum is at the center of the SW pointing which has the largest continuum fraction and smallest beam-scaling exponent of the three pointings.  The continuum peak is within the CEN pointing and its fraction is almost equal to the SW but with a higher index.  The NE pointing's continuum fraction is the smallest because it is off the peak.

\begin{deluxetable}{cccc}
\tablecaption{Continuum Fit Results\label{tab:contfit}}
\tablewidth{0pt}
\tablehead{
\colhead{Pointing} &\multicolumn{2}{c}{Continuum Fit} & \colhead{Reduced $\chi^{2}$}\\
 & \colhead{$A$} & \colhead{$B$} & 
 }
\startdata
NE     & 0.318 $\pm$ 0.001 & 1.03 $\pm$ 0.03 & 2.4 \\
CEN  & 0.413 $\pm$ 0.001 & 1.05 $\pm$ 0.02 & 3.2 \\
SW    & 0.425 $\pm$ 0.001 & 0.99 $\pm$ 0.02 & 3.7
\enddata
\tablecomments{The continuum fit columns give the coupling fraction ($A$) and beam scaling exponent ($B$) of the fit described in Section \ref{ContFit} and defined in equation \ref{eqn:ContFit}.  The reduced $\chi^{2}$ values are for the line and continuum fits and are calculated based solely on statistical errors and do not include errors due to calibration.  The fits use all data points except the lowest five channels in each spectrum and have 132 degrees of freedom.}
\end{deluxetable}

\subsection{Spectral Line Fitting Results}

The channel-to-channel spacing in Z-Spec is 500--1300~MHz, corresponding to 700--1200 km s$^{-1}$, thus the instrument does not resolve the line profiles in M82.   Nevertheless, it is possible to fit integrated intensities and center frequencies by comparison with the instrumental response for each bolometer, carefully measured in the laboratory with a long-path Fourier-transform spectrometer.   Each spectral line (indexed by $j$) is modeled as a Gaussian profile of center frequency $\nu_j$, amplitude $A_j$, and FHWM $\delta\nu_j$.  The sum of these line profiles plus the continuum is multiplied by the (normalized) measured spectral profile of each channel used in the fitting (indexed by $i$) $f_i(\nu)$ and integrated over the range of the measured profiles: 180--320~GHz.  This process creates a model Z-Spec spectrum, which can be compared to the observed spectrum.  The input line frequencies and amplitudes are then varied to arrive at a least-squares  fit, using statistical ($1/\sigma_i^2$) weighting.  The frequencies of the known lines are fixed in the fitting, modulo a common redshift, so that the fit determines the redshift but not the line frequencies.  In practice for local-Universe galaxies, the $^{12}$CO transition dominates the redshift determination.  With this method, we obtain accurate centroid measurements of line features that did not have an obvious identification.  The fitted line intensities are given in Table \ref{tab:lineinten} using the adopted $\Delta v=250$ km s$^{-1}$ for all lines in the three pointings; reducing the FWHM to 100 km s$^{-1}$ produces nearly identical integrated intensities and fit quality.  We comment here on our spectral line measurements and how they compare to those found in the literature.

\begin{deluxetable*}{ccccccc}
\tabletypesize{\scriptsize}
\tablecaption{M82 Line Fit Results\label{tab:lineinten}}
\tablewidth{0pt}
\tablehead{
\colhead{Species \&}  & \colhead{Rest Freq.} & \colhead{$E_{\mathrm{upper}}$} & 
\colhead{Beam Size} &
\multicolumn{3}{c}{Integrated Line Intensity (K km s$^{-1}$)} \\
\colhead{Transition} & \colhead{(GHz)} & \colhead{(K)} & \colhead{(arcsec)} & 
\colhead{NE} & \colhead{CEN} & \colhead{SW}
}
\startdata
       CO~\jtwo       & 230.538 & 16.6 & 32 & 423.9 $\pm$ 3.0 & 460.0 $\pm$ 3.2 & 452.5 $\pm$ 3.3 \\
$^{13}$CO~\jtwo       & 220.399 & 15.9 & 34 & 29.4 $\pm$ 0.7 & 31.7 $\pm$ 0.7 & 30.2 $\pm$ 0.7 \\
C$^{18}$O~\jtwo       & 219.560 & 15.8 & 34 & 8.6 $\pm$ 0.7 & 7.9 $\pm$ 0.7 & 7.8 $\pm$ 0.7 \\
       CN~\jtwo        & 226.875 & 16.3 & 33 & 14.3 $\pm$ 0.6 & 23.3 $\pm$ 0.8 & 19.9 $\pm$ 0.6 \\
       CS~\jfour      & 195.954 & 23.5 & 37 & 4.3 $\pm$ 0.8 & 4.4 $\pm$ 0.8 & 4.1 $\pm$ 0.7 \\
       CS~\jfive      & 244.936 & 35.3 & 31 & $<$ 1.6 [1.2] & 2.3 $\pm$ 0.5 & $<$ 1.5 [1.1] \\
       CS~\jsix       & 293.912 & 49.4 & 26 & $<$ 3.1 [0.8] & $<$ 3.3 [3.0] & $<$ 3.1 [1.5] \\
\\
     \hco~\jthree     & 267.558 & 25.7 & 29 & 13.7 $\pm$ 0.8 & 16.9 $\pm$ 0.9 & 15.4 $\pm$ 0.9 \\
      HCN~\jthree     & 265.886 & 25.5 & 29 & 5.1 $\pm$ 0.8 & 7.0 $\pm$ 0.8 & 6.9 $\pm$ 0.8 \\
      HNC~\jthree     & 271.981 & 26.1 & 28 & 2.4 $\pm$ 0.6 & 2.7 $\pm$ 0.7 & 3.5 $\pm$ 0.6 \\
     \cch~\jthree        & 262.251 & 25.2 & 29 & 7.9 $\pm$ 0.7 & 11.1 $\pm$ 0.7 & 11.0 $\pm$ 0.6 \\
\\
    \hhco~$(3_{13}\too 2_{12})$   & 211.211 & 32.1 & 35 & 4.0 $\pm$ 0.6 & 2.6 $\pm$ 0.6 & $<$ 1.6 [1.5] \\
    \hhco~$(3_{03}\too 2_{02})$*  & 218.222 & 21.0 & 34 & 2.3 $\pm$ 0.6 & 3.6 $\pm$ 0.6 & 3.1 $\pm$ 0.6 \\
    \hhco~$(4_{14}\too 3_{13})$   & 281.527 & 45.6 & 27 & $<$ 2.5 [1.7] & $<$ 2.6 [1.6] & 3.3 $\pm$ 0.9 \\
    \hhco~$(4_{04}\too 3_{03})$*  & 290.623 & 34.9 & 27 & $<$ 3.1 [1.0] & 3.7 $\pm$ 1.0 & 3.6 $\pm$ 1.0 \\
    \hhco~$(4_{13}\too 3_{12})$   & 300.837 & 47.9 & 26 & 4.0 $\pm$ 1.1 & $<$ 3.5 [2.6] & 4.0 $\pm$ 1.1 \\
\\
   \chcch~\jtwelve    & 205.081 & 64.0 & 36 & 5.1 $\pm$ 0.6 & 4.3 $\pm$ 0.6 & 4.2 $\pm$ 0.6 \\
   \chcch~\jthirteen  & 222.167 & 74.6 & 33 & 2.5 $\pm$ 0.7 & 3.4 $\pm$ 0.7 & 2.5 $\pm$ 0.7 \\
   \chcch~\jfourteen  & 239.252 & 86.1 & 31 & $<$ 1.9 [1.0] & 2.2 $\pm$ 0.6 & 2.1 $\pm$ 0.6 \\
   \chcch~\jfifteen   & 256.337 & 98.4 & 30 & $<$ 2.2 [1.0] & $<$ 2.2 [1.4] & 2.3 $\pm$ 0.7
\enddata
\tablecomments{Measured line intensities and their estimated uncertainties or the three-$\sigma$ upper limits for the 20 identified transitions in the three pointings.  For for the upper limits, the value is three times the 1-$\sigma$ uncertainty in the fitted intensity and the fitted intensity value is given in square brackets.  The adopted linewidth for all measurements is 250 km s$^{-1}$ (see text); changing this value by factors of two in either direction does not significantly change the fitted line intensities or the quality of the fit.  The errors are based on the statistical errors in the spectral data and do not include any uncertainty due to calibration.  The hyperfine splitting in the CN and \cch transitions is not accessible at Z-Spec's resolution.  The \hhco lines marked with an * may be blended with other higher energy \hhco transitions.}
\end{deluxetable*}

\subsubsection{CO and its Isotopologues}
The CO, $^{13}$CO, and C$^{18}$O~\jtwo\ transitions have been extensively studied and mapped by previous experiments \citep{Mao:2000fr,Weis:2001zr}; however, direct comparison with published intensities is limited by beam size mismatch.  \citet{Wild:1992qy} quote integrated main-beam intensities nearly twice what we measure in a 13\arcsec\ beam for the CO~\jtwo\ transition, implying a beamsize ($\theta$) scaling of $I\propto\theta^{-0.8}$ whereas the main species intensities given in \citet{Mao:2000fr} in a 22\arcsec\ beam indicate a beam scaling exponent of $-1.2$ to $-1.5$.  The maps obtained by \citet{Thuma:2000uq} show that the CO emission is more extended than the continuum emission which would indicate the exponent should be slightly less than unity.  The isotopologues $^{13}$CO and C$^{18}$O have been measured in \citet{Mao:2000fr} and \citet{Wild:1992qy}, respectively, and though they indicate different beam scaling exponents as they did for the main species, they both imply that the isotopologue distribution is slightly more concentrated than the main species.  

\subsubsection[CN and \cch]{CN and C$_2$H}
We detect both the CN~\jtwo\ and \cch~\jthree\ transitions in all pointings.  The primary energy levels of both of these molecules are split by hyperfine interactions, giving spectra with more structure than is accessible using Z-Spec.  The intensities quoted in Table \ref{tab:lineinten} are obtained by fitting a single Gaussian profile and thus represent the total integrated intensity for all transitions.  These species have been detected, for the first time, in M82 by \citet{Henkel:1988mz} and \citet{Fuente:2005ly}, but the \jthree\ transition of  \cch is a new detection.  Neither of these authors quotes the intensity seen in all hyperfine components, making a direct comparison difficult.  Our emphasis is on a study of physical conditions which, if it included these species, would require knowledge of the intensities of the hyperfine components.  Therefore, we do not include the CN and \cch in this analysis that follows.

\subsubsection{HCO$^+$, HCN, and HNC}
The \jthree\ transitions of the HCO$^+$, HCN, and HNC molecules are strongly detected in all three pointings.  The \hco and HCN transitions have been previously detected by \citet{Wild:1992qy}, however, the only previous detection of HNC in M82 has been the \jone\ line by \citet{Huettemeister:1995uq}.  Comparing the 12\arcsec\ beam measurements from \citet{Wild:1992qy} to ours suggest a beam scaling exponent of roughly $-1$, consistent with the continuum and CO values.  We use our measurements and measurements of other transitions in the subsequent analysis.

\subsubsection{CS}
The \jfour, $5\too4$, and $6\too5$ transitions of CS lie in the Z-Spec band and it's simple ladder of rotational transitions make it an ideal candidate for study with our instrument.  Unfortunately, the lines in our band are not very bright in M82 so we can only give upper limits for the \jsix\ lines and the \jfive\ lines in the NE and SW pointings.  The transitions we detect have been seen in M82 by \citet{Bayet:2008sf} for the CEN pointing and \citet{Bayet:2009fu} for the NE and SW pointings.  Longer integration times should enable the first detections of the \jsix\ lines, particularly in the CEN pointing where the fitted intensity is 2.7$\sigma$.  As with \hco, HCN, and HNC, we combine our measurements with the other measured transitions for the analysis that follows.

\subsubsection{H$_2$CO}
We have identified the $3_{03}\too 2_{02}$, $3_{13}\too 2_{12}$, $4_{14}\too 3_{13}$, $4_{04}\too 3_{03}$, and $4_{13}\too 3_{12}$ transitions of formaldehyde (\hhco) which are detected in at least one of the three pointings; the latter four of these transitions have not been detected previously.   Two of these transitions, $3_{03}\too 2_{02}$ and $4_{04}\too 3_{03}$, may be blended with other higher energy \hhco lines and our reported integrated intensity should be interpreted as the sum of the intensities of all of these blended lines.  The feature at 218 GHz has been studied with the heterodyne receiver array HERA on the IRAM 30m telescope by \citet{Muhle:2007rr}.  They detected three formaldehyde transitions, $3_{03}\too 2_{02}$, $3_{22}\too 2_{21}$, and $3_{21}\too 2_{20}$, near our NE and SW pointings along with a possible detection of a methanol line in the NE pointing.  In addition, they demonstrated the power of using formaldehyde lines to trace both temperature and density of the molecular gas.  Higher-resolution followup of the lines we have identified would provide a powerful extension to their work.

\subsubsection{CH$_3$CCH}
Methyl acetylene (\chcch) has the largest number of atoms of any molecule detected outside our galaxy and several transitions have been seen previously in the SW lobe of M82 \citep{Mauersberger:1991fj}.  Our measurements in the NE and CEN pointings as well as all the measurements of the \jtwelve\ and \jfifteen\ transitions are new.  We suspect that the \jfifteen\ transition may be contaminated with an unidentified feature that is causing the fitted flux in the SW pointing to be unrealistically high.  This spectral feature is right at our three-sigma threshold and additional data are needed to precisely determine the line identifications. 

\section{Analysis}
\label{sec:analysis}

\subsection{Excitation and Radiative Transfer Modeling} \label{sec:rad}

We turn now to a study of the physical conditions in the dense gas, as probed with rotational transitions of HCN, HNC, HCO$^+$, and CS.   To make useful inferences about the conditions in M82's molecular gas, modeling is required.  The approach is to adopt basic input parameters such as total amount of gas in the beam (column density), gas density, temperature, and abundance of the species under consideration and then calculate the resulting line intensities.  A grid of such calculations over ranges of input parameters can then provide a framework to interpret the observations.  We note that the transitions we study have a range of critical densities ranging from 10$^4$ to 10$^8$ cm$^{-3}$.

We use the RADEX code \citep{VanderTak:2008} for our excitation and radiative transfer modeling.  The primary inputs to RADEX are the choice of molecule, the kinetic temperature of the molecular gas $T_{\mathrm{kin}}$, the density of molecular hydrogen $n_{\mathrm{H_{2}}}$ in cm$^{-3}$, and the column density of the species $N_{\mathrm{mol}}$ in cm$^{-2}$.  The radiative transfer calculation depends on $N_{\mathrm{mol}}/\Delta v$, where $\Delta v$ is the velocity width of the line.  Z-Spec cannot measure the linewidth, so $\Delta v = 250$ km s$^{-1}$ is used throughout the radiative transfer modeling to be consistent with the linewidth used for the spectral fits.  
RADEX does not assume local thermal equilibrium, but uses an escape probability formalism that connects the optical depth to the chance an emitted photon escapes the source cloud.  Several different physical models for this escape probability have been derived; we use the expanding spherical shell model.  However, the results are very insensitive to the choice of escape probability.  Starting with an initial guess for the level population distribution, RADEX computes the optical depths of all the molecule's transitions, from which a new level population distribution can be calculated.  This process iterates until a self-consistent solution is achieved such that the optical depth changes by less than a default tolerance from one iteration to the next.    We use the collisional excitation rates calculated in 
\citet{Lique:2006qy} and \citet{Lique:2007uq} for the CS species and the rates from the online database outlined in \citet{Schoier:2005ve} for the three remaining species.

\subsection{Parameter Likelihood Estimation} \label{sec:likecalc}

Calculated line intensities are then compared to the intensity measurements of the species under consideration.  An additional area filling factor parameter $\Phi_{\mathrm{A}}$ must be included in the model because the clumps of gas producing the radiation do not in general fill the beam.  $\Phi_{\mathrm{A}}$ scales down the line intensities from RADEX so that they can be directly related to the measurements.  When $\Phi_{\mathrm{A}}$ is less than 1, $N_{\mathrm{mol}}$ represents the column density of an individual radiating clump while the product $\Phi_{\mathrm{A}}N_{\mathrm{mol}}$ is the beam-averaged column density, $<\!\!N_{\mathrm{mol}}\!\!>$.  

The line-intensity measurements are inherently uncertain and the physical interpretation using models should reflect that.  A Bayesian method for calculating likelihood distributions for various physical quantities of interest can be used to address this measurement uncertainty (\citealp{Ward:2002ul,Ward:2003ai}\defcitealias{Ward:2003ai}{W03}[hereafter \citetalias{Ward:2003ai}]).  The method constructs the probability distribution of obtaining the measurements with their associated errors given a set of physical parameters, assuming the measurements are independent and the errors are Gaussian distributed.  Using a prior-probability density function for the range of physical parameters, the probability distribution of the measurements given the physical parameters can be inverted into a likelihood distribution for the physical parameters given the measurement results.

Bayes' Postulate says that the prior probability density function should be uniform for all cases in the absence of prior knowledge.  The prior probability distributions used for this analysis are assumed to be logarithmically uniform in all model parameters.  However, this prior probability is used to exclude certain non-physical situations relating to large column densities and small molecular hydrogen densities.  Both constraints require knowledge of the molecular abundance ratio, $X_{\mathrm{mol}} \equiv n_{\mathrm{mol}}/n_{\mathrm{H_{2}}}$.  These constraints, described in detail by \citetalias{Ward:2003ai}, limit the total molecular mass contained in the telescope beam to less than the dynamical mass of the galaxy ($2.0\times 10^9 \, M_{\odot}$, based on the estimates given by \citetalias{Ward:2003ai} and \citet{Panuzzo:2010}\defcitealias{Panuzzo:2010}{P10}[hereafter \citetalias{Panuzzo:2010}]) and limit the column length, equal to the column density divided by the number density, to less than the length of the bright molecular emission on the plane of the sky.  In addition, models with optical depths larger than 100 in any transition are excluded because very large optical depths are not appropriate for the species under consideration and RADEX is not accurate when the optical depth is this large.

A critical aspect of the analysis is the scaling of the published measurements of various transitions to a common beamsize.  The limits of beam scaling are  $\theta^{0}$ for a source that fills the beam for all measurements, and $\theta^{-2}$ for a source that is always smaller than the beam.  M82 is in an intermediate range with respect to the 25\arcsec -- 35\arcsec\ Z-Spec beam; it is neither fully point-like nor beam-filling.  Based on the CO and mm-wave continuum maps, we use an intermediate beam scaling of $\theta^{-1}$, appropriate for the distribution which is to first order extended along the major-axis, but unresolved along the minor axis.

\subsection{Multi-Species Model}

\begin{deluxetable*}{cccccccl}
\tabletypesize{\scriptsize}
\tablecaption{High-Dipole-Moment Species Measurements\label{tab:rtlmeas}}
\tablewidth{0pt}
\tablehead{
\colhead{Transition} & \colhead{Rest Freq.} & \colhead{$E_{\mathrm{upper}}$} & 
\colhead{Obs. Beam} & \colhead{NE Flux} & \colhead{CEN Flux} & \colhead{SW Flux} &
\colhead{Refs} \\
 & \colhead{(GHz)} & \colhead{(K)} & \colhead{(arcsec)} & 
 \colhead{(K km s$^{-1}$)} & \colhead{(K km s$^{-1}$)} & \colhead{(K km s$^{-1}$)} &
}
\startdata
\multicolumn{8}{c}{CS} \\
\jone   &  48.991 &  2.4 & 36 & ... & 16.2 $\pm$ 1.1 & ... & 2 \\
\jtwo   &  97.981 &  7.1 & 25.1 &  9.4 $\pm$ 0.2 & 13.3 $\pm$ 0.3 &  8.9 $\pm$ 0.2 & 3, 4 \\
\jthree & 146.969 & 14.1 & 16.7 &  8.9 $\pm$ 0.1 & 11.2 $\pm$ 0.3 &  7.6 $\pm$ 1.7 & 3, 4, 5 \\
\jfour  & 195.954 & 23.5 & 37.1 &  4.3 $\pm$ 0.8 &  4.4 $\pm$ 0.8 &  4.1 $\pm$ 0.8 & 1 \\
\jfive  & 244.936 & 35.3 & 30.8 &  $<$ 0.5 & 2.3 $\pm$ 0.5 & $<$ 0.5 & 1 \\
\jsix   & 293.912 & 49.4 & 26.3 & $<$ 1.0 & $<$ 1.1 & $<$ 1.0 & 1 \\
\tableline
\multicolumn{8}{c}{C$^{34}$S} \\
\jthree & 144.617 & 13.9 & 17 & $\sim$0.5 $\pm$ 50\% &  0.6 $\pm$ 0.1 &  $\sim$0.4 $\pm$ 50\% & 6 \\
\jfour  & 192.818 & 23.1 & 37.1 & $<$ 0.8 & $<$ 0.8 & $<$ 0.8 & 1 \\
\jfive  & 241.016 & 34.7 & 30.8 & $<$ 0.5 & $<$ 0.5 & $<$ 0.5 & 1 \\
\jsix   & 289.209 & 48.6 & 26.3 & $<$ 1.0 & $<$ 1.1 & $<$ 1.0 & 1 \\
\tableline
\multicolumn{8}{c}{\hco} \\
\jone   &  89.189 &  4.3 & 23 & 35.0 $\pm$ 2.0 & 38.3 $\pm$ 2.0 & 37.2 $\pm$ 2.0 & 7 \\
\jthree & 267.558 & 25.7 & 28.6 & 13.7 $\pm$ 0.8 & 16.9 $\pm$ 0.9 & 15.4 $\pm$ 0.9 & 1 \\
\jfour  & 356.734 & 42.8 & 14 & 23.6 $\pm$ 1.7 & 22.2 $\pm$ 1.7 & 22.6 $\pm$ 1.9 & 8 \\
\tableline
\multicolumn{8}{c}{HCN} \\
\jone   &  88.632 &  4.3 & 23 & 21.8 $\pm$ 2.0 & 18.5 $\pm$ 2.0 & 23.7 $\pm$ 2.0 & 7 \\
\jthree & 265.886 & 25.5 & 28.8 &  5.1 $\pm$ 0.8 &  7.0 $\pm$ 0.8 &  6.9 $\pm$ 0.8 & 1 \\
\jfour  & 354.505 & 42.5 & 14 &  5.6 $\pm$ 0.6 &  9.0 $\pm$ 0.7 &  6.1 $\pm$ 0.4 & 8 \\
\tableline
\multicolumn{8}{c}{HNC} \\
\jone   &  90.664 &  4.4 & 25 & 10.7 $\pm$ 2.0 & 13.4 $\pm$ 0.8 & 12.3 $\pm$ 3.0 & 9 \\
\jthree & 271.981 & 26.1 & 28.2 &  2.4 $\pm$ 0.6 &  2.7 $\pm$ 0.7 &  3.5 $\pm$ 0.6 & 1
\enddata
\tablecomments{Fluxes and upper limits used in our multi-species radiative transfer likelihood analysis.  In addition to the given statistical error, a 10\% calibration error is added in quadrature to each measurements' uncertainty.  An additional 10\% error is added to the lines measured in beams smaller than 18\arcsec\ and to the CS \jone\ line.  The CS \jone\ line has only been measured for the CEN pointing and thus is not included for the likelihood analysis for the other pointings.  The C$^{34}$S \jthree\ has also only been measured in the CEN pointing; we estimate the flux in the other pointings based on the \jthree\ line ratio between CS and C$^{34}$S in the CEN pointing and apply a 50\% error to these estimates for the likelihood analysis.  The C$^{34}$S upper limits are based on the uncertainty in the spectral fit for the main-species lines; the actual integrated intensity in the C$^{34}$S lines is well below the detection threshold of our data.}
\tablerefs{
(1) This Work; 
(2) \citealt{Paglione:1995dk};
(3) \citealt{Bayet:2008sf};  
(4) \citealt{Bayet:2009fu};  
(5) \citealt{Mauersberger:1989wd};  
(6) \citealt{Martin:2009ve};  
(7) \citealt{Nguyen-Q-Rieu:1989pd};   
(8) \citealt{Seaquist:2000qf};  
(9) \citealt{Huettemeister:1995uq}. 
}
\end{deluxetable*}

Four of the species detected in this survey have the required radiative and collisional rate data available in an online database \citep{Schoier:2005ve} or in the literature \citep{Lique:2006qy,Lique:2007uq}: \hco, HCN, HNC, and CS.  The method described in \citetalias{Ward:2003ai} and outlined in the previous section was developed for analyzing 11 transitions of CO and  $^{13}$CO and it can be applied to each of the four molecules above individually.  However, for three of the four selected molecules, the number of detected transitions (see Table \ref{tab:rtlmeas}) is less than the four primary parameters of the radiative transfer model ($T_{\mathrm{kin}}$, $n_{\mathrm{H_{2}}}$, $N_{\mathrm{mol}}$, and $\Phi_{\mathrm{A}}$).  CS and C$^{34}$S have had several transitions measured but with relatively low signal-to-noise which would lead to very broad constraints from the likelihood analysis.  

Observations of these four species in star-forming regions within our Galaxy indicate good correspondence with both the spatial distributions and the line profiles \citep{Brand:2001yf,Nikolic:2003it}.  High spatial resolution maps of the Galactic circumnuclear disk \citep{Christopher:2005wq} and of the starburst galaxy NGC 253 \citep{Knudsen:2007ec} in \hco and HCN indicate strong similarity, in general, between the emission of these two molecules.  

We thus construct a model in which all four species are characterized by a common kinetic temperature, molecular hydrogen density, molecular hydrogen column density ($N_{\mathrm{H_{2}}}$), and filling factor.   Each species is modeled with a individual abundance ($X_{\mathrm{mol}}$).  Extraction of the absolute abundances relative to H$_2$ is not possible with RADEX, but it can constrain the relative abundances of the various species.   In the following analysis, CS is chosen as the primary species and the model is parameterized by $T_{\mathrm{kin}}$, $n_{\mathrm{H_{2}}}$, $N_{\mathrm{CS}}$, $\Phi_{\mathrm{A}}$, $X_{\mathrm{HCO}^+}/X_{\mathrm{CS}}$, $X_{\mathrm{HCN}}/X_{\mathrm{CS}}$, $X_{\mathrm{HNC}}/X_{\mathrm{CS}}$, and $X_{\mathrm{C^{34}S}}/X_{\mathrm{CS}}$.

\begin{figure}
\begin{center}
\includegraphics[width=\columnwidth]{\fullpath{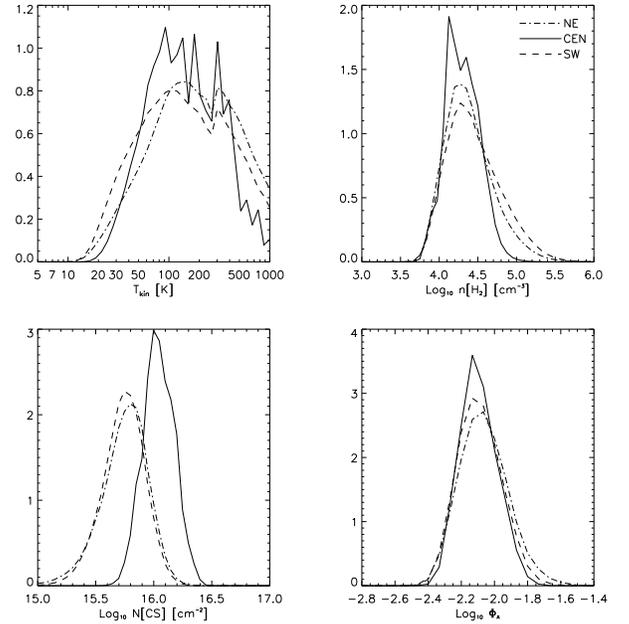}}
\caption[Primary Parameter Likelihood Distributions]{Likelihood distributions for the four primary species parameters, $T_{\mathrm{kin}}$, $n_{\mathrm{H_{2}}}$, $N_{\mathrm{CS}}$, and $\Phi_{\mathrm{A}}$, for the five-species radiative transfer likelihood analysis.  All of the distributions have well defined peaks, indicating the parameters are well constrained by the measurements within the range of the RADEX simulation.  The distributions for the three pointings are plotted in dash-dotted, solid and dashed lines for the NE, CEN, and SW pointings, respectively.  The distributions are normalized to have unit integral when integrated over the base-10 logarithm of the parameter value.  The jaggedness seen in the $T_{\mathrm{kin}}$ distributions is caused by two factors: first, the spike seen in the distributions for all three pointings at 300 K is caused by patching together the two sets of colisional rates of CS from \citet{Lique:2006qy} and \citet{Lique:2007uq}.  The former paper calculated the rates for 31 rotational levels for temperatures up to 300 K while the later produced rates for 38 rotational and 3 vibrational levels in a higher temperature range.  The remaining jaggedness in the CEN pointing $T_{\mathrm{kin}}$ distribution probably comes from a numerical problem in RADEX where, for certain physical conditions, it fails to converge on a stable solution for the optical depth in the CS \jone\ line.  The NE and SW pointings' distributions do not show this effect because they do not have a CS \jone\ measurement.
\label{fig:rtlprim1D}}
\end{center}
\end{figure}

\begin{figure}
\begin{center}
\includegraphics[width=\columnwidth]{\fullpath{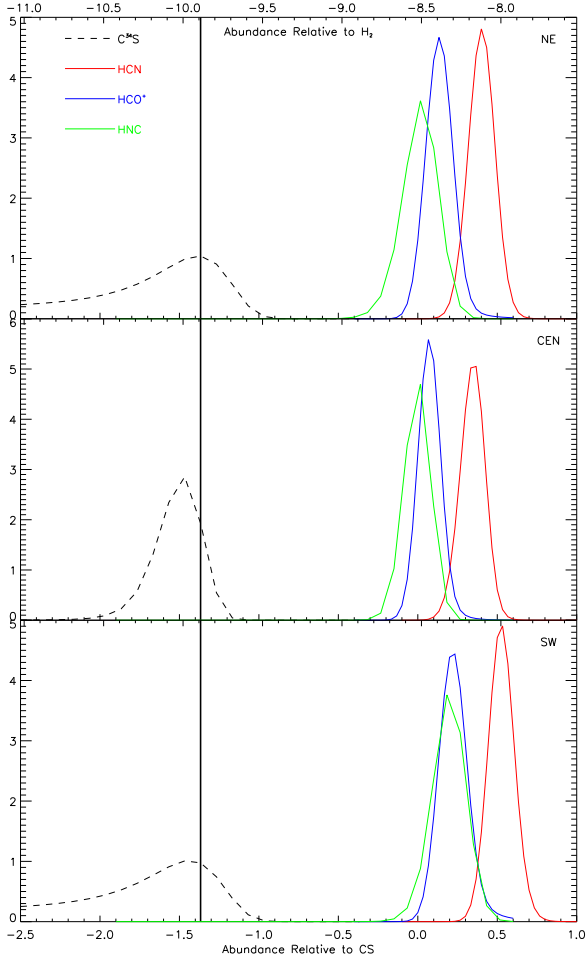}}
\caption[Secondary Species Likelihood Distributions]{Likelihood distributions for the abundances of the secondary species, \hco, HCN, HNC, and C$^{34}$S in blue, red, green, and dashed black, respectively, for the NE (top), CEN (middle), and SW (bottom) pointings.  The thick vertical black line indicates the solar abundance of $^{34}$S/$^{32}$S $= 0.043$ that is assumed for calculating all other likelihood distributions.  The bottom axis is labeled with the abundance relative to CS, which is the parameterization used in the model.  The top axis shows the abundance relative to H$_2$ using the adopted value of $X_{\mathrm{CS}} = 3\times 10^{-9}$.
\label{fig:rtlsec}}
\end{center}
\end{figure}

\begin{figure}
\begin{center}
\includegraphics[width=\columnwidth]{\fullpath{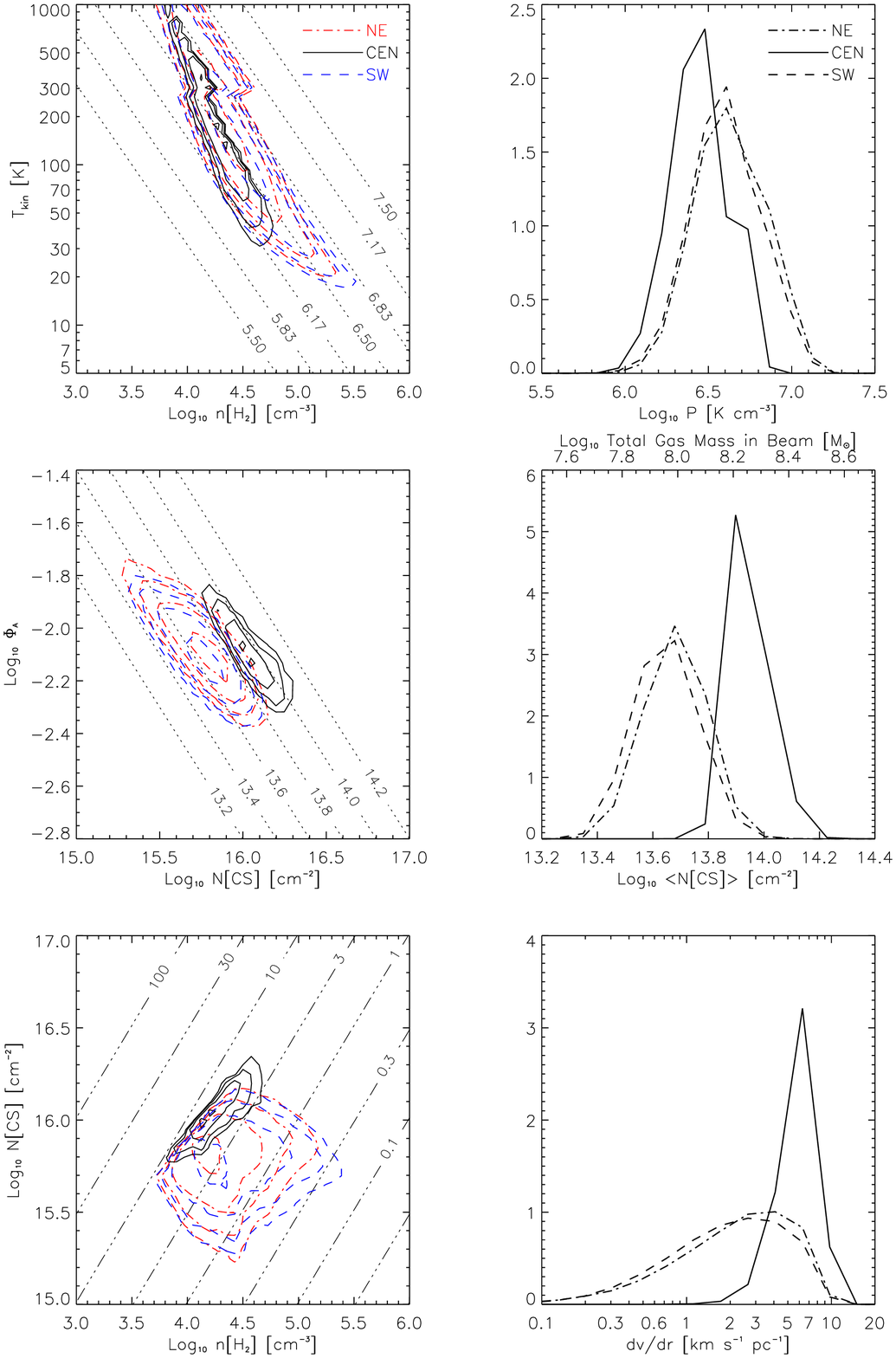}}
\caption[Secondary Parameter Likelihood Distributions]{The plots in the left column show two-dimensional likelihood distributions for three pairs of primary species parameters while the right column plots show distributions for parameters that are projected from the corresponding 2D distributions.  The three pointings, NE, CEN, and SW, are indicated by red, black, and blue contour lines in the 2D distributions and with dash-dotted, solid, and dashed lines in the projected 1D distributions.  The contour lines represent 10\%, 20\%, 40\%, and 80\% of the peak of the 2D likelihood distributions.  The top-left plot has dashed lines of constant pressure, labeled in units of $\log_{10}$ K cm$^{-3}$; the $T_{\mathrm{kin}}$ versus $n_{\mathrm{H_{2}}}$ distributions are used to create the pressure distributions shown in the top-right plot.  Dashed lines of constant beam-averaged column density in the left-middle $\Phi_{\mathrm{A}}$ versus $N_{\mathrm{CS}}$ plot show how the $<\!\!N_{\mathrm{CS}}\!\!>$ distributions in the right-middle plot are obtained.  The alternative axis in the beam-averaged column density plot shows the molecular mass in the beam assuming $X_{\mathrm{CS}} = 3\times 10^{-9}$. The bottom-left plot shows the $N_{\mathrm{CS}}$ versus $n_{\mathrm{H_{2}}}$ distributions; the diagonal dash-dot-dot-dot lines indicate constant velocity gradient in units of km s$^{-1}$ pc$^{-1}$ at the most-likely value of $\Phi_{\mathrm{A}}$.  The velocity gradient distributions shown in the bottom-right panel are computed from the three-dimensional distributions of $N_{\mathrm{CS}}$, $n_{\mathrm{H_{2}}}$, and $\Phi_{\mathrm{A}}$.
\label{fig:rtlprim2D}}
\end{center}
\end{figure}

The result of the likelihood analysis is a likelihood matrix with each point in the matrix characterized by a particular value of the four primary species parameters and three secondary species abundance ratios which are used to parameterize the model.  As described in \citetalias{Ward:2003ai}, likelihood distributions for a single parameter can be obtained by integrating the likelihood matrix along all the other dimensions.  These distributions, seen in Figures \ref{fig:rtlprim1D} and \ref{fig:rtlsec}, can be used to calculate both median values and confidence ranges for the seven parameters in our model (see Table \ref{tab:MultModResults}).  It is also possible to calculate likelihood distributions for parameters which are functions of model parameters, such as the gas pressure, $P = n_{\mathrm{H_{2}}} \times T_{\mathrm{kin}}$, and beam-averaged column density, $<\!\!N_{\mathrm{CS}}\!\!> = \Phi_{\mathrm{A}} \times N_{\mathrm{CS}}$.  The beam-averaged column density can be used to calculate the total molecular mass in the beam by
\begin{equation}
\label{eqn:massinbeam}
M_{\mathrm{beam}} = 1.5\times m_{H_{2}} \times \frac{\pi D^2_{\mathrm{beam}}}{4} \times \frac{<\!\!N_{\mathrm{mol}}\!\!>}{X_{\mathrm{mol}}}
\end{equation}
where $m_{H_{2}}$ is the mass of a hydrogen molecule, $D_{\mathrm{beam}}$ is the linear diameter of the beam in cm and the factor of 1.5 accounts for the additional mass of He and dust in the molecular clouds.  Distributions for these parameters and their associated two-dimensional distributions are shown in Figure \ref{fig:rtlprim2D}.  

\begin{deluxetable*}{ccccccc}
\tabletypesize{\scriptsize}
\tablecaption{Multi-Species Modeling Results\label{tab:MultModResults}}
\tablewidth{0pt}
\tablehead{
\colhead{Quantity} & \multicolumn{2}{c}{NE Pointing} &  \multicolumn{2}{c}{CEN Pointing} & 
 \multicolumn{2}{c}{SW Pointing} \\
 & \colhead{Median} & \colhead{Range} & \colhead{Median} & \colhead{Range} & 
 \colhead{Median} & \colhead{Range}
}
\startdata
 & \multicolumn{6}{c}{Primary Species Parameters} \\
$T_{\mathrm{kin}}$ (K) & 160 & 58 - 470 & 130 & 56 - 320 & 130 & 44 - 410 \\
$n_{\mathrm{H_{2}}}$ (cm$^{-3}$) & 10$^{4.3}$ & 10$^{4.0}$ - 10$^{4.7}$ & 10$^{4.2}$ & 10$^{4.0}$ - 10$^{4.5}$ & 10$^{4.4}$ & 10$^{4.1}$ - 10$^{4.8}$ \\
$N_{\mathrm{CS}}$ (cm$^{-2}$) & 10$^{15.7}$ & 10$^{15.5}$ - 10$^{15.9}$ & 10$^{16.0}$ & 10$^{15.9}$ - 10$^{16.2}$ & 10$^{15.7}$ & 10$^{15.5}$ - 10$^{15.9}$ \\
$\Phi_{\mathrm{A}}$ & 10$^{-2.1}$ & 10$^{-2.3}$ - 10$^{-1.9}$ & 10$^{-2.1}$ & 10$^{-2.3}$ - 10$^{-2.0}$ & 10$^{-2.1}$ & 10$^{-2.3}$ - 10$^{-2.0}$ \\
\tableline
 & \multicolumn{6}{c}{Secondary Species Relative Abundances} \\
$X_{\mathrm{HCO^{+}}}/X_{\mathrm{CS}}$ & 10$^{0.1}$ & 10$^{0.0}$ - 10$^{0.2}$ & 10$^{0.06}$ & 10$^{-0.02}$ - 10$^{0.13}$ & 10$^{0.2}$ & 10$^{0.1}$ - 10$^{0.3}$ \\
$X_{\mathrm{HCN}}/X_{\mathrm{CS}}$ & 10$^{0.4}$ & 10$^{0.3}$ - 10$^{0.5}$ & 10$^{0.3}$ & 10$^{0.2}$ - 10$^{0.4}$ & 10$^{0.5}$ & 10$^{0.4}$ - 10$^{0.6}$ \\
$X_{\mathrm{HNC}}/X_{\mathrm{CS}}$ & 10$^{0.0}$ & 10$^{-0.2}$ - 10$^{0.1}$ & 10$^{0.0}$ & 10$^{-0.2}$ - 10$^{0.1}$ & 10$^{0.2}$ & 10$^{0.0}$ - 10$^{0.3}$ \\
$X_{\mathrm{C^{34}S}}/X_{\mathrm{CS}}$ & 10$^{-1.7}$ & 10$^{-2.7}$ - 10$^{-1.3}$ & 10$^{-1.6}$ & 10$^{-1.7}$ - 10$^{-1.4}$ & 10$^{-1.8}$ & 10$^{-2.7}$ - 10$^{-1.4}$ \\
\tableline
 & \multicolumn{6}{c}{Projected Parameters} \\
Pressure (K cm$^{-3}$) & 10$^{6.6}$ & 10$^{6.3}$ - 10$^{6.8}$ & 10$^{6.4}$ & 10$^{6.2}$ - 10$^{6.6}$ & 10$^{6.5}$ & 10$^{6.3}$ - 10$^{6.8}$ \\
$<\!\!N_{\mathrm{CS}}\!\!>$ (cm$^{-2}$) & 10$^{13.6}$ & 10$^{13.5}$ - 10$^{13.8}$ & 10$^{13.9}$ & 10$^{13.8}$ - 10$^{14.0}$ & 10$^{13.6}$ & 10$^{13.4}$ - 10$^{13.7}$ \\
$\mathrm{d}v / \mathrm{d}r$ (km s$^{-1}$ pc$^{-1}$) & 1.9 & 0.6 - 4.4 & 4.5 & 3.0 - 7.2 & 1.6 & 0.5 - 4.1 \\
Total Gas Mass in Beam ($M_\odot$) & 10$^{7.9}$ & 10$^{7.8}$ - 10$^{8.1}$ & 10$^{8.2}$ & 10$^{8.1}$ - 10$^{8.3}$ & 10$^{7.9}$ & 10$^{7.8}$ - 10$^{8.0}$
\enddata
\tablecomments{Results obtained from the multiple species radiative transfer modeling of the lines of CS, \hco, HCN, HNC and C$^{34}$S.  The median and 68\% ($1\sigma$) confidence ranges are obtained from the likelihood distributions shown in Figures \ref{fig:rtlprim1D} -- \ref{fig:rtlprim2D}.}
\end{deluxetable*}

The emitting regions containing these molecules are likely to be isolated cores of high-density gas.  In that limit, we can estimate the velocity gradient inside the cores with a simple geometrical argument.  The characteristic size of an individual core can be estimated by $S_{\mathrm{core}} \approx (N_{\mathrm{CS}}/X_{\mathrm{CS}})/n_{\mathrm{H_{2}}}$, which is simply the length defined by the ratio of the molecular hydrogen column and volume densities.  This value can be used to estimate the number of cores in the beam by computing the ratio of the area of the emitting region in the beam over the size of a single core, $N_{\mathrm{core}} \approx (\Phi_{\mathrm{A}} \times D^2_{\mathrm{beam}})/S^2_{\mathrm{core}}$.  A reasonable approximation for isolated cores where $\Phi_{\mathrm{A}} \ll 1$ is that the observed total line width $\Delta v$ is split up equally among the individual cores.  That implies that the velocity gradient in a single core is
\begin{equation}
\frac{\mathrm{d}v}{\mathrm{d}r} \approx \frac{\Delta v/N_{\mathrm{core}}}{S_{\mathrm{core}}}
= \frac{\Delta v}{D^2_{\mathrm{beam}}X_{\mathrm{CS}}}\times
\frac{N_{\mathrm{CS}}}{\Phi_{\mathrm{A}}n_{\mathrm{H_{2}}}}.
\end{equation}
The distributions for the core velocity gradient are shown in Figure \ref{fig:rtlprim2D} along with  two-dimensional projections of the three-dimensional distributions used for the calculation.

\subsection{Molecular Abundances}

\begin{figure*}
\begin{center}
\includegraphics[width= \textwidth]{\fullpath{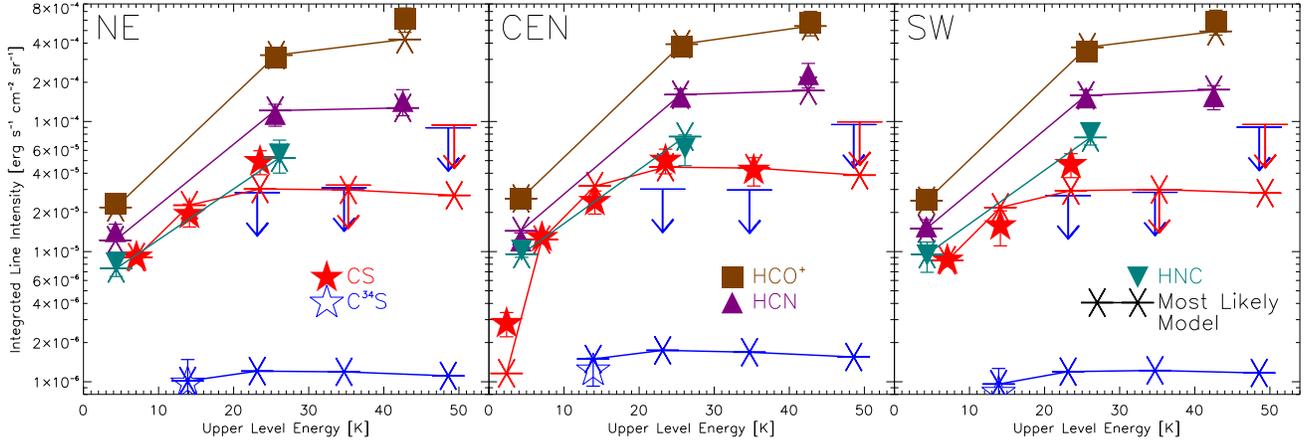}}
\caption[Measured and Modeled Multi-Species SED]{Measured and modeled spectral energy distributions for the five molecules of the multi-species radiative transfer likelihood analysis for the three pointings, NE (left), CEN (center), and SW(right).  The measurements of CS, \hco, HCN, HNC, and C$^{34}$S, from Table \ref{tab:rtlmeas} scaled to a common 30\arcsec\ beam, are marked with red stars, brown squares, purple triangles, teal triangles and open blue stars, respectively.  The error bars include both measurement and adopted calibration error, if any.  Three-sigma upper limits are indicated with downward arrows using the same color scheme as the measurements.  The most-likely model from the likelihood analysis is indicated by the six-pointed stars connected by lines, in colors to match the measurements.
\label{fig:measmodSED}}
\end{center}
\end{figure*}

The likelihood distributions of the \hco, HCN, HNC, and C$^{34}$S abundances relative to that of CS are shown in Figure \ref{fig:rtlsec}.  These represent the first statistically rigorous measurements of molecular abundances in M82.  Average values for the entire starburst nucleus are presented in Table \ref{tab:molabund}.  As discussed in Section \ref{sec:likecalc}, the abundance of CS is used in the likelihood calculation to apply certain physical limits on the parameter space of the radiative transfer grid.  We adopt the CS abundance $X_{\mathrm{CS}}$ of $3\times 10^{-9}$ calculated in \citet{Mauersberger:1989wd}.  Their calculation is based on assuming optically thin CS emission and the CO intensity to H$_2$ column density conversion factor that is observed in the Milky Way.  Changing the CS abundance by half an order of magnitude either up or down does not affect the likelihood distributions for nearly all of the parameters; the molecular hydrogen density and kinetic temperature distributions shift down and up, respectively, with increasing CS abundance such that the distribution of gas pressure is relatively unchanged.  The agreement between the measured and most-likely model's integrated line intensities, shown graphically in Figure \ref{fig:measmodSED} for $X_{\mathrm{CS}} = 3\times 10^{-9}$, is not significantly impacted by changing $X_{\mathrm{CS}}$.  Furthermore, while the likelihood distributions for the C$^{34}$S abundance show slight differences between the three pointings, we have little reason to suspect that this value would be much different from the solar isotopic abundance ratio of $^{34}$S/$^{32}$S $= 0.043$.  Therefore, we impose this isotopic ratio for the other likelihood calculations.

\begin{deluxetable}{cccccc}
\tablecaption{Molecular Abundance Ratios for M82\label{tab:molabund}}
\tablewidth{0pt}
\tablehead{
\colhead{Species} & \colhead{CS}  & \colhead{\hco} & \colhead{HCN} & \colhead{HNC} & 
\colhead{C$^{34}$S}
}
\startdata
$\log_{10}(X_{\mathrm{mol}})$  & -8.5 &  -8.4 & -8.1   & -8.5   &  -9.9
\enddata
\tablecomments{Abundance ratios (with respect to H$_{2}$) derived from the multiple species modeling results for the abundance of \hco, HCN, HNC, and C$^{34}$S relative to CS and the abundance for CS relative to H$_2$ from \citet{Mauersberger:1989wd}.  The uncertainty in the four modeled abundances is $\pm 0.1$ and is less than $\pm 0.5$ for CS in logarithmic units.}
\end{deluxetable}

Of course, changing the assumed CS abundance impacts the conversion from the relative abundance ratios to standard abundances relative to H$_2$.  Also, while the beam-averaged column density distribution does not change when using different values for the CS abundance, the conversion from $<\!\!N_{\mathrm{CS}}\!\!>$ to total mass is inversely proportional to the abundance (see equation \ref{eqn:massinbeam}).  Of the three values of CS abundance we have used, the value from \citet{Mauersberger:1989wd} seems to be the most reasonable; lowering it drives the total molecular mass too high and raising it increases the temperatures to extreme values. 

While our likelihood analysis is an independent confirmation of the CS abundance estimated by \citet{Mauersberger:1989wd}, it should be noted that there is significant debate about the CO intensity ($I_{\mathrm{CO}}$) to molecular hydrogen column density ($N_{\mathrm{H_{2}}}$) conversion factor, $\mathcal{X}_{\mathrm{CO}} \equiv N_{\mathrm{H_{2}}}/I_{\mathrm{CO}}$, used in their calculation of CS abundance.  Their value was 2.2 - 2.5 times larger than more recent measurements of $\mathcal{X}_{\mathrm{CO}}$ in the Milky Way ($1.6 \times 10^{20}$ cm$^{-2}$ (K km s$^{-1}$)$^{-1}$, \citealp{Hunter:1997fk}; $1.8 \times 10^{20}$ cm$^{-2}$ (K km s$^{-1}$)$^{-1}$, \citealp{Dame:2001uq}).  Furthermore, a detailed analysis of an interferometric CO \jone\ map of the nucleus of M82 showed that $\mathcal{X}_{\mathrm{CO}}$ is roughly 2-10 times less than what is measured in the Milky Way \citep{Weis:2001zr}.  Quantitative modeling of ensembles of unresolved giant molecular clouds by \citet{Maloney:1988kx} showed that $\mathcal{X}_{\mathrm{CO}}$ is proportional to $T_{\mathrm{kin}}^{-1}\,n_{\mathrm{H_{2}}}^{1/2}$; this proportionality was confirmed in M82 by \citet{Weis:2001zr}.  Studies of several galaxies have also shown that $\mathcal{X}_{\mathrm{CO}}$ also depends on metalicity \citep{Wilson:1995vn, Boselli:2002ys} and CO intensity \citep{Nakai:1995zr}.  The deviations from the standard Milky Way value can be significant; \citet{Casasola:2007ly} measured the conversion factor in several giant molecular associations in M81, the interaction partner of M82 and NGC 3077 \citep{Yun:1994fr}, and found the galactic average of $\mathcal{X}_{\mathrm{CO}}$ to be 20 times larger than the Milky way value.  In spite of these developments in the understanding of $\mathcal{X}_{\mathrm{CO}}$ since the publication of \citet{Mauersberger:1989wd}, our results indicate that their estimate for the CS abundance is correct to within a half an order of magnitude.

Adopting the value $X_{\mathrm{CS}} = 3\times 10^{-9}$ allows for comparison of our abundance measurements to those found in the literature.  For the most part, M82 is assumed to have abundances similar to those found in regions of high-mass star formation in our Galaxy such as Sgr B2 or Orion.  \citet{Wild:1992qy} quote $X_{\mathrm{HCO^{+}}} = 1\times 10^{-8}$ and $X_{\mathrm{HCN}} = 2\times 10^{-8}$ which have also been used by other authors \citep[e.g.,][]{Seaquist:2000qf}.  These are somewhat higher than what we measure though we agree that HCN is twice as abundant as \hco.  \citet{Huettemeister:1995uq} use their measurements of HCN \jone\ and the CO intensity to H$_2$ column density conversion factor to calculate $X_{\mathrm{HNC}} = 1.4\times 10^{-10}$, assuming the HCN emission is optically thin.  Our measurements of the HNC \jthree\ contradict the optically-thin assumption and our models point to an abundance 1.5 orders of magnitude larger, roughly equal to the CS abundance.

\subsection{Mass of Dense Gas}
The total mass of molecular gas in the nuclear region can be estimated from the total mass in the beam measured for each of the three pointings by adopting a particular geometry for the central region.  If the area of the nucleus is $A_\mathrm{n}$ and the area of overlap between our 30\arcsec\xspace beam and this nuclear area is $A_\mathrm{ol}$, then the total mass in the nucleus can be calculated using
\begin{eqnarray}
\label{eqn:totalmassingal}
M_{\mathrm{total}} & = & M_{\mathrm{CEN}}
	+M_{\mathrm{NE}}\left(\frac{A_\mathrm{n} - 
						A_\mathrm{ol}}{2A_\mathrm{ol}}\right) \nonumber\\
& & +M_{\mathrm{SW}}\left(\frac{A_\mathrm{n} - A_\mathrm{ol}}{2A_\mathrm{ol}}\right),
\end{eqnarray}
where $M_{\mathrm{CEN}}$, $M_{\mathrm{NE}}$, and $M_{\mathrm{SW}}$ are the masses measured in the three pointings.  This equation uses the mass surface density implied by $M_{\mathrm{NE}}$ and $M_{\mathrm{SW}}$ to account for the mass not covered on the left and right sides of the CEN pointing.  If we adopt a rectangular nuclear geometry 50\arcsec\xspace $\times$ 15\arcsec\xspace, as indicated by the CO \jtwo\ interferometer maps in \citet{Weis:2001zr}, then the area scaling factor $(A_\mathrm{n} - A_\mathrm{ol})/2A_\mathrm{ol} = 0.37$ and $M_{\mathrm{total}} = 2.2 \pm 0.5 \times 10^8 \, M_{\odot}$ in the nuclear starburst.   This value is comparable to the total mass traced in CO from both multi-line studies (\citetalias{Ward:2003ai}; 2.0$\times 10^{8} \, M_\odot$ when corrected to $d=3.86$ Mpc) and via interferometric low-J CO and $^{13}$CO imaging \citep{Weis:2001zr}.   

Since we measure a gas mass with a suite of transitions, we can provide a calibration of the HCN $\mathcal{X}$-factor, $\mathcal{X}_{\rm HCN}$, which converts HCN line luminosity (in temperature units) to mass of dense gas.   Assuming that the source couples to the various beam sizes as $\theta^1$, we find $\mathcal{X}_{\mathrm{HCN,} J=1}$ of 10--15, in agreement with the canonical value of 10 derived from virial considerations by \citet{Gao:2004b,Gao:2004a}.   Since HCN is sub-thermally excited, using the \jthree\ transition $\mathcal{X}_{\rm HCN}$ is much higher with values of 36--65. 

\section{Discussion}

\subsection{Physical Conditions and Relationship to CO-Traced and Atomic Gas}
Our likelihoods suggest temperatures between 50--500~K, broadly consistent with the warm components modeled by \citetalias{Ward:2003ai}, up to \jsix, and somewhat lower than the 400--800~K derived by \citetalias{Panuzzo:2010} in considering all of the transitions up to \jthirteen\ as measured with {\it Herschel} SPIRE.  Our results are also consistent with the temperature of 200~K inferred from the formaldehyde measurements and analysis of \citet{Muhle:2007rr}. The relatively low precision with which we measure the temperature is not surprising since the most likely temperatures are generally higher than the upper level energies of the transitions we are studying (e.g., HCN $J=3$:  $T=25$ K).  We do note that our results do not support the presence of substantial amounts of gas at temperatures below $\sim$30~K unless the density is so low that the transitions in our analysis would not be excited.   This means that the drop in line intensity with $J$ (in temperature units) is due to sub-thermal excitation of the levels above $J=1$, and the run of line intensity with $J$ should thus provide a reliable density measurement.   Indeed, our derived median densities are 1.5--3 $\times 10^4\,\rm cm^{-3}$, sub-critical for the transitions above 200 GHz.   

Our derived densities are larger than those derived with CO studies. \citetalias{Ward:2003ai} find 600--6000~cm$^{-3}$ and \citetalias{Panuzzo:2010} find 1000--13000, though we do note in some cases \citetalias{Ward:2003ai} find densities poorly constrained on the high-density end.   The product of temperature and density is the thermal pressure, and we find values of 1--4$\times10^{6}\,\rm K\,cm^{-3}$, an order of magnitude higher than the pressure inferred by \citetalias{Ward:2003ai} (0.5--4$\times10^{5}\,\rm K\,cm^{-3}$), but comparable to that derived by \citetalias{Panuzzo:2010} using all of the CO transitions.   Our high densities may reflect the fact that we are probing preferentially high-density cores, in approximate pressure equilibrium with the larger, more diffuse envelopes which produce the bulk of the CO.  This should not be surprising since HCN, HNC, and CS are generally found in UV-shielded cores as their dissociation energies are less than that of CO, and they don't have generally achieve sufficient column densities to self-shield.  For instance, the photo-dissociation region (PDR) chemical models of \citet{Fuente:2008} show that HCN exists primarily within $A_\mathrm{V}>5$.

We compare our results with the studies of the photo-dissociated atomic gas.  \citet{Kaufman:1999} and \citet{Colbert:1999dq} have applied a PDR model \citep[updated from][]{Wolfire:1990,Tielens:1985}\footnote{see also \tt{http://dustem.astro.umd.edu/pdrt/index.html}} to [C {\small II}] and [O {\small I}] fine-structure-line measurements from the Kuiper Airborne Observatory (KAO) and ISO Long-wavelength Spectrometer (LWS), respectively.  PDR conditions are parametrized in terms of the density and UV field strength $G_0$.  Estimates for the M82 central starburst range from: 1) $n_{\mathrm{H}_2}$=$10^4\,\rm cm^{-3}$,  $G_0$=$10^{3.5}$ (\citealp{Kaufman:1999}, assuming the [C {\small II}] emission is uniformly distributed over its 55\arcsec\ beam, so only a small fraction arises in the [O {\small I}]-emitting region),  2) $n_{\mathrm{H}_2}$=$10^{2.7}\,\rm cm^{-3}$,  $G_0 = 10^{2.5}$ (\citealp{Kaufman:1999}, assuming that all the large-beam [C {\small II}] emission arises in the same region as the [O {\small I}]), and 3) $n_{\mathrm{H}_2}$=$10^{3.3}\,\rm cm^{-3}$, $G_0 = 10^{2.8}$ (\citealp{Colbert:1999dq}, using large-beam ISO fluxes, but removing a [C {\small II}] contribution from the ionized gas).   These estimates form a locus in the $n_{\mathrm{H}_2}$, $G_0$ plane, with the upper end of the density range becoming consistent with our density likelihood.  The UV-illuminated surfaces of clouds might be expected to have somewhat lower density than the UV-shielded cores, and we note that with the modeled surface temperatures of $\sim$300~K, the PDR thermal pressures range from 10$^{5.3}$ to 10$^{6.5}$, broadly consistent with the values derived from the CO as well as our analysis.

However, the PDR models which fit the atomic line fluxes cannot explain the strength of the mid-$J$ CO transitions.   For the range of PDR conditions inferred from the atomic lines, the modeled CO \jseven\ to [C {\small II}] intensity ratio is at most $\sim1.7\times10^{-3}$ (at $n$=10$^4\,\rm cm^{-3}$).  The CO spectrum indeed peaks (in energy units) at \jseven\ per the {\it Herschel} SPIRE measurements \citepalias{Panuzzo:2010}.  This transition carries a fraction 1.4--2.3\% of the [C {\small II}], depending on whether the [C {\small II}] is resolved or unresolved in its 55\arcsec\ beam, relative to the 43\arcsec\ CO analysis region.   Thus the CO emission in M82 exceeds the PDR predictions for the measured densities by an order of magnitude.  Moreover, this is a lower limit---if the [C {\small II}] includes a contribution from ionized gas \citep[as assumed by][]{Colbert:1999dq}, then the inferred PDR line ratio is even larger, and less consistent with the model.   Higher-density PDR models $n>10^{4.5}\,\rm cm^{-2}$ can bring the mid-$J$ CO to [C {\small II}] into agreement with the observations, but then the [O {\small I}] transition is over-predicted relative to the observations by a factors of 3--10.  Moreover, such high-densities for the PDR would be inconsistent with our measured density likelihood which excludes $n>10^{4.5}\,\rm cm^{-3}$.  

Empirically, the suite of [C {\small II}], [O {\small I}], and mid-$J$ CO emission in M82 does not compare with the PDRs associated with Galactic star-formation regions.   In M17, the bright mid-$J$ CO and its widespread distribution with [C {\small II}] over several parsecs as viewed in the edge-on PDR led  \citet{Stutzki:1988,Harris:1987} to a clumpy PDR interpretation.   However, their data indicate that the CO \jseven\ in this source only amounts to some 0.5--1$\times10^{-3}$ relative to the [C {\small II}], much less than in M82.   The Orion PDR is more extreme, with CO \jseven\ some 8\% of the [C {\small II}] \citep{Schmidt-Burgk:1990,Stacey:1993}, but in this case the [O {\small I}] 63~\mm\ line dominates the PDR line emission with 11$\times$ the [C {\small II}] (see \citealp{Herrmann:1997}), so the CO \jseven\ is $\sim$0.8\% of the atomic line emission.  While the partitioning of energy between the atomic and warm molecular components in the Orion bar is thus similar to what we observe in M82,  we stress again that the line ratio are not a good match, particularly the [O {\small I}] to [C {\small II}], which is only 1--1.5 in M82.  The bright CO and [O {\small I}] in Orion is consistent with a clumpy PDR model which includes clumps with density as high as 10$^7\,\rm cm^{-3}$ \citep{Burton:1990,Koester:1994, Meixner:1993}.   Such densities cannot be commonplace in M82 based on line ratios in the atomic gas, the CO analyses, and our analysis of the high-dipole-moment species which indicate typical densities less than 10$^{4.5}\,\rm cm^{-3}$.

\subsection{Heating of the Gas}

The poor match to the Galactic PDRs and the inability of the PDR models to explain the powerful mid-$J$ CO emission suggests that non-UV heating sources may be dominating the energetics of the molecular material in this starburst nucleus.  X-rays can be a powerful source of energy input to the gas, producing luminous X-ray Dissociation Regions \citep[XDRs,][]{Maloney:1996}. However, the hard X-ray luminosity of M82 is only $1.1\times 10^6 L_\odot$ \citep{Strickland:2007}, completely inadequate to power the observed CO emission.  Moreover, multiple chemical / excitation studies show that the line emission from M82 is not consistent with X-rays being a dominant heating term.   The XDR models of  \citet{Meijerink:2005} and \citet{Meijerink:2007} predict more [O {\small I}] than is observed \citep{Colbert:1999dq} at our measured densities around 10$^4\,\rm cm^{-3}$.   \citet{Fuente:2008} studied HOC$^+$ and HCO$^+$ in detail and showed that the HOC$^+$ emission is not spatially correlated with X-ray emission, and the line ratios among these species as well as CO$^+$ and CN are well-matched with PDR models with $n>10^4\,\rm cm^{-3}$ and $G_0 = 10^4$, similar to the values derived in analysis of the atomic lines.  They did not consider the total energy budget of the molecular gas, and did not discuss heating mechanisms.

Simiarly, \citet{Loenen:2008} have considered XDR models and PDR models with and without extra bulk molecular heating and have predicted line ratios among the HCN, HCO$^+$, and HNC \jone\ transitions.  According to their models, the measurements for M82 (Table \ref{tab:rtlmeas}) are inconsistent with XDRs, as XDRs are predicted to have HNC comparable to or brighter than HCN.   The line ratios are best fit with PDRs with $n\sim10^{4.5}\,\rm cm^{-3}$ (their lowest density considered), but with additional bulk heating on the order of $3\times10^{-19}\,\rm erg\,s^{-1}cm^{-3}$, $\sim 1\, L_\odot/M_\odot$ input into the molecular material (they ascribe this to mechanical heating).  

We thus look for other potential bulk heating sources other than X-rays.  \citet{Suchkov:1993} proposed that the elevated level of cosmic rays due to the supernova rate in M82 will heat the gas, and their derived cosmic ray enhancement factor relative to the Galaxy of $\sim$500 has been confirmed by recent measurements of high-energy gamma-rays in M82 \citep{Veritas_M82:2009}.   \citet{Bradford:2003} showed that the energy input from a similar cosmic ray density is indeed well-matched to the total molecular gas cooling in NGC~253 as extrapolated from the CO transitions up to  \jseven, where the CO emission appears to peak, \citep{HaileyDunsheath:2008}.  Because the CO excitation and total cooling in M82 are similar (from \citetalias{Ward:2003ai}, \citetalias{Panuzzo:2010} fluxes), we conclude that cosmic rays are a plausible means of heating the gas.

Another important heating term for bulk molecular material is the dissipation of turbulence \citep{Falgarone:1995, MacLow:1999, Pan:2009, Bradford:2005}.   The average heating per gram is given by dimensional analysis $0.5 \times \sqrt{3} \sigma_v^3 / L$, where $L$ is the driving scale for the turbulence and $\sigma_v$ is the (1-D) RMS velocity spread on this size scale.  \citet{Pan:2009} conclude that turbulent dissipation with $\sigma_v\sim2.5\,\rm km\,s^{-1}$ on 0.1--1~pc scales is the dominant heating term on average in Galactic clouds (exceeding cosmic ray heating by factors of 3--4), and that it produces temperatures of 13--36 K in Galactic clouds with densities of 10$^4$--10$^5$ cm$^{-3}$.  For gas densities above $10^6$ cm$^{-3}$, gas-grain cooling becomes dominant, and limits the increase in the gas temperature, as the dust energy balance is largely independent of the gas temperature.  A lower bound to the total cooling per mass in M82 is given by the total CO luminosity in \citetalias{Panuzzo:2010} ratioed to the mass derived by \citetalias{Ward:2003ai} using the mid-$J$ lines, $\sim$0.1~$L_\odot/M_\odot$.  Achieving this via turbulent dissipation requires $\sigma_v\sim$5--20$\,\rm km\,s^{-1}$ on the same 0.1--1~pc scales.  The velocity gradient implied by our radiative transfer analysis of 4--10~$\rm km\,s^{-1}\,pc^{-1}$ is a bit lower than this would suggest, but it becomes consistent if the turbulence is distributed on few-pc scales (e.g., $\sigma_v\sim20\,\rm km\,s^{-1}$, $L\sim$5 ~pc), as might be expected if it is produced by winds and supernova shocks from young star clusters.

\subsection{Comparison with Expanding Shell Starburst Models}

Our findings are largely consistent with the evolving starburst model of \citet{Yao:2006} and \citet{Yao:2009}, in which the gas in the nucleus of M82 is a superposition of expanding spherical bubbles around stellar clusters.   The bubble interiors are ionized gas ([H {\small II}] regions), and the shells are swept-up molecular gas, so the inside edges are PDR fronts.  The PDR surfaces are responsible for most of the atomic line emission, and would agree with the \citet{Fuente:2008} results.  The PDR shell also generates most of the excited CO emission, though their model includes mechanical energy input from the shock due to the expansion, which may be a substantial term in the heating of the warm molecular gas.   

Model intensities for the high-dipole moment species are not presented, preventing a detailed comparison with our data, but (not surprisingly) the total mass and physical conditions in their modeled shells are similar to what we find with our likelihood analysis.  Their estimated gas mass of $2\times10^8\,M_\odot$ from CO in the central 1~kpc along the major axis (nearly identical to our modeled 50\arcsec\xspace $\times$ 15\arcsec\xspace region) is comparable to our measured $2.2\times10^8\,M_\odot$, and the shell density at the putative 3--10~Myr age is modeled to be 1--3$\times10^4\,\rm cm^{-3}$, similar to what we find.   However, as \citet{Yao:2009} notes, there are some inconsistencies in the model.  While all of the observed molecular and atomic line emission is reproduced, the stellar luminosity which is required is only $\sim$5\% of the observed far-IR luminosity in the same region, potentially the result of assuming zero pressure for the ambient ISM which results in more mass swept up in the modeled shells than is physical. 

\subsection{Warm Star-Forming Gas}

Regardless of the details, the \citet{Yao:2009} model is representative of the likely physical situation: a new stellar cluster subjects the surrounding molecular gas to both UV photon and mechanical energy input which heats and compresses it, at least in the first 10~Myr after the starburst.   The result is a molecular ISM that is demonstrably warmer than the Galactic cloud cores.  Does this mean that the SF is quenched?   \citet{Fuente:2008} compare line ratios of ions (HOC$^+$, CO$^+$ to HCN) with their PDR chemistry model to estimate the total depth of the PDR (ions except HCO$^+$ quickly become less abundant with increasing $A_{\rm V}$).  Fitting the line ratios to two components, they find some $\sim$87\% of the molecular gas is in small clouds with $A_{\rm V}\sim5$ (but large enough to house HCN, HNC) with only $\sim13\%$ in clouds with $A_{\rm V}\sim50$ and conclude that in general the molecular gas is highly fragmented with clouds too small to form massive stars.  \citet{Forster:2003} found that the SF in M82 has occurred in two bursts, one in the center some 10~Myr ago and one in a circumnuclear ring $\sim$5~Myr ago, and that each burst was self-quenching with a timescale of a few Myr due to mechanical energy input into the gas.   The total mass of stars formed through both episodes is modeled at 2--5$\times 10^{8}\,M_\odot$, depending on the low-mass part of the IMF (and cannot be more than $\sim6\times 10^8\,M_\odot$, the total measured stellar mass in the system).  The stellar mass formed in the last 10 Myr is thus comparable to or at most double the amount of dense molecular gas remaining, so unless the eventual star formation efficiency is limited to 30--50\%, one may ask if the gas can be the raw material for another round of SF.

If the material we trace is indeed forming stars, then the warm molecular medium is likely to impact the stellar IMF, increasing the fraction of high-mass stars by inhibiting the formation of low-mass stars.  Theoretical studies of the IMF all involve scaling from a Jeans mass, the mass at which a cloud's self-gravity overcomes its support forces \citep[e.g.,][]{Larson:2005}.  The support can be either simple thermal pressure or large-scale turbulent motions.  In their recent analytical study, \citet{Hennebelle:2008} note that for typical ISM physical conditions and a reasonable prescription for the turbulence, the turbulent support is more important for the high-mass end of the spectrum, while the evolution of lower-mass condensations are governed by simple thermal support.

 The thermal Jeans mass can be written as $M_\mathrm{J} = 1.1 M_\odot (T/10 K)^{1.5} \rho_{19}^{-0.5}$ (where $\rho_{19}$ is the mass density in units of $10^{-19}\,\rm g\,cm^{-3}$) and yields 50~$M_\odot$ for our median derived temperature (120~K) and density (10$^{4.2}\,\rm cm^{-3}$).  This may be indicating that the bulk of the material is indeed unlikely to participate meaningfully in any further SF in its present condition.   Of course, the SF will occur in the densest and coolest regions, but they would likely be in approximate pressure equilibrium with the bulk of the gas.  If we consider the lowest temperatures allowed by our likelihood analysis, $T\sim$30~K, together with the highest pressures, log $P\sim$ 6.7, then the density is $n=10^{5.2}\, \rm cm^{-3}$, and the Jeans mass is ${M_\mathrm{J}}\sim3 \, M_\odot$.\footnote{We note for completeness that there is evidence for a (mass-independent) efficiency factor that relates the mass of a Jeans-unstable core to the mass of the actual star which forms from it, believed to be $\sim$1.4--2 \citep{Hennebelle:2008}, meaning that the resulting stellar masses are somewhat smaller than the Jeans mass estimates.}    
 
 A meaningful comparison with the Galaxy is hampered by the fact that our large-beam M82 observations are necessarily averaging over multiple SF regions, and will include gas in outflows as well as collapsing protostars themselves.  Our approach is to examine the material around the Galaxy's most massive SF sites, since they are likely the best Galactic examples of SF on large scales.   We consider the sample studied by \citet{Leurini:2007} in the millimeter and centimeter-band methanol transitions which are used to derive accurate temperatures and densities.  We consider only the envelopes rather than the cores since the cores appear to be heated internally and are presumably already undergoing collapse, and in any case, the envelopes dominate the mass of these regions.   
 
 \citet{Leurini:2007} find temperatures ranging from 11--36~K, and densities of 10$^5$--10$^6\,\rm cm^{-3}$.  Thus even these massive star formation sites are cooler on average than the dense gas in M82.   The lower temperature is not surprising; again, in dense regions the gas temperature will approach the dust temperature, which in the Galaxy ranges from 10--20~K \citep{Paradis:2009}.  While the inferred thermal pressures in these star forming envelopes are comparable to those we find in M82, the lower temperature and higher density corresponds to a smaller typical Jeans mass---values range from 0.3--2.8 $M_\odot$, less than the minimum $\sim 3 \,M_\odot$ derived above for M82.    
 
More generally, the characteristic formed stellar mass scale $M_*$ is seen to scale as $T_\mathrm{min}^\gamma$, where $T_\mathrm{min}$ is the minimum temperature to which the gas can cool, and the exponent $\gamma$ ranges from 1.7 \citep[obtained in numerical experiments,][]{Jappsen:2005} to 3.35 \citep[via an analytic treatment,][]{Larson:1985}.  If we take the measured minimum of $\sim$30~K versus a conservative 20~K in the warm Galactic regions, this scaling suggests a factor of at least 2--4 in $M_*$ for M82 relative to the Galaxy.  Clearly, accurate estimates require theoretical study and more detailed knowledge of the local conditions at the star formation sites, but if the gas we are tracing is indeed involved in star formation, then it  likely produces a stellar IMF which is biased against low-mass stars relative to even the massive star formation sites in the Galaxy.
  
Such a scenario is of course consistent with the reports of low-mass-deficient stellar populations in M82 over the years \citep{Rieke:1980,Rieke:1993,Forster:2003}.  Moreover, an IMF biased against low-mass stars produces more luminosity per unit stellar mass than if the IMF is as observed in the Galaxy.   Such a top-heavy or bottom-light IMF has been proposed to explain an apparent discrepancy between the observed stellar mass buildup and the energy release history in the first half of the Universe ($z>1$) \citep{Perez-Gonzalez:2008, Dave:2008}.   Given that the typical star forming galaxy in this epoch is now believed to be similar to the local LIRGs and ULIRGs \citep{LeFloch:2005, Papovich:2007}, the conditions in M82 are likely more indicative of the historical average than those of the Galaxy.

\section{Conclusions}
We present a study of the dense molecular gas in the starburst nucleus of M82 based on \zspecrange spectra toward three positions obtained with the Z-Spec instrument.   Z-Spec offers good sensitivity, accurate continuum measurement, and a uniform calibration for spectral lines across this band.  We report fluxes for some 20 molecular transitions, many new detections.   The measurements of the \jthree\ transitions of HCO$^+$, HCN, and HNC, and the \jfour\ and \jfive\ transitions of CS motivate an excitation and radiative transfer analysis in which all four species are simultaneously considered, incorporating all of their available published transitions.  Our analysis constrains the physical conditions in the dense gas as well as the relative abundances among these species.   We trace some 1.7--2.7$\times10^{8} \,M_\odot$ of gas with $n_{\mathrm{H_{2}}}\simeq$1--3$\times10^4\,\rm cm^{-3}$, and find that it is warm: likely above 50~K and potentially as high as 500~K, a range which exceeds the level temperature of the transitions studied.  The mass and temperature are thus comparable to that found for the warm component in the mid-$J$ CO studies, but the higher density implies a thermal pressure of 1.5--4$\times10^6\,\rm K\, cm^{-3}$, about an order of magnitude higher then inferred from the mid-$J$ CO transitions.  

In the framework of physical and chemical models, the line ratios among HCN, HCO$^+$, and HNC indicate that the molecular gas is subject to both UV photons as well as a bulk heating mechanism other than X-rays.  A similar conclusion is reached in considering the direct observed cooling in the CO lines up to \jseven.   Cosmic ray heating and dissipation of mechanical energy from the new star clusters are both potential heating sources for the molecular ISM in M82.  This feedback has rendered much of the molecular ISM in the nucleus sterile to further SF.  We briefly compare the dense molecular gas in M82 with star-forming sites in the Galaxy, concluding that if any of the material we are studying is involved in further SF, then the increased heating likely biases the stellar IMF against low-mass stars, relative to the Galaxy.  Such a scenario may be more indicative of the typical SF environment in the Universe's history than the Galactic stellar IMF.

\acknowledgments

We are deeply grateful to the staff of the Caltech Submillimeter Observatory for their help in Z-Spec's commissioning and observing.  We acknowledge Peter Ade and his group for some of our filters and Lionel Duband for the $^3$He/$^4$He refrigerator in Z-Spec and are thankful for their help in the early integration of the instrument.  We also appreciate the comments and careful reading of an anonymous referee.  Finally, we acknowledge the following grants and fellowships: NSF CSO grant (AST-0838261) for B.~Naylor, NASA SARA grants NAGS-11911 and NAGS-12788, an NSF Career grant (AST-0239270) and a Research Corporation Award (RI0928) to J.~Glenn, a Caltech Millikan and JPL Director's fellowships to C~M.~Bradford, an NSF grant (AST-0807990) and an NRAO Jansky fellowship to J.~Aguirre, and NASA GSRP fellowships to L.~Earle and J.~Kamenetzky.

The research described in this paper was carried out at the Jet Propulsion Laboratory, California Institute of Technology, under a contract with the National Aeronautics and Space Administration. \copyright\ 2010.  All rights reserved. 

\bibliography{m82paper_apj}

\end{document}